\documentclass[11pt]{article}
\usepackage{a4wide}
\usepackage{amssymb, amsmath, amsthm}
\usepackage{multirow}
\usepackage{natbib}
\usepackage[dvips]{graphicx}

\usepackage{fancyhdr}
\DeclareMathOperator*{\argmax}{arg\,max}

\renewcommand{\l}{\ell}
\renewcommand{\H}{\mathcal{H}}
\newcommand{\A}{\mathcal{A}}
\newcommand{\R}{\mathcal{R}}
\newcommand{\ind}[1]{{\mathbf{1}\{#1\}}}
\newcommand{\sbullet}{1:n}

\title{Spatial clustering of array CGH features in combination with hierarchical multiple testing}

\author{Kyung In Kim, Etienne Roquain and Mark A. van de Wiel}

\date{}

\date{National Cancer Institute, UPMC University of Paris 6 and VU University - Medical Center}

\begin{document}

\maketitle
\begin{abstract}
We propose a new approach for clustering DNA features using array CGH data from multiple tumor samples. We distinguish data-collapsing: joining contiguous DNA clones or probes with extremely similar data into regions, from clustering: joining contiguous, correlated regions based on a maximum likelihood principle. The model-based clustering algorithm accounts for the apparent spatial patterns in the data. We evaluate the randomness of the clustering result by a cluster stability score in combination with cross-validation. Moreover, we argue that the clustering really captures spatial genomic dependency by showing that coincidental clustering of independent regions is very unlikely.
Using the region and cluster information, we combine testing of these for association with a clinical variable in an hierarchical multiple testing approach. This allows for interpreting the significance of both regions and clusters while controlling the Family-Wise Error Rate simultaneously. We prove that in the context of permutation tests and permutation-invariant clusters it is allowed to perform clustering and testing on the same data set. Our procedures are illustrated on two cancer data sets.
\end{abstract}

\section*{Introduction}
Array Comparative Genomic Hybridization (array CGH) was designed as a high-resolution measurement device for copy number aberrations, which are known to be involved in the cancer development process \citep{Kallioniemi2008}. Through hybridization, fluorescent labeled DNA is extracted from test and reference samples resulting in log-ratios of intensities of the two types of samples. When plotted against chromosomal position, the log-ratio data appears as segments at various amplitudes. One is particularly interested in those chromosomal segments which show levels of loss or gain, because those segments may possibly harbor oncogenic genes, cancer progression markers or other relevant features. Array CGH data is often encoded as loss (deletion), normal and gain by assigning discrete copy number states to chromosomal segments. Generally, two processing steps are performed for this purpose: segmentation and calling.

Segmentation denotes a process to find breakpoints from single copy number profiles. Many segmentation algorithms have been proposed. \cite{Willenbrock2005, lai05bioinfo} discuss and compare several of those, while \cite{DR07} provide software to simultaneously apply several popular algorithms to the same data set. Recently, new algorithms have been introduced to deal with the computational burden caused by the increasing resolution of the arrays \cite[f.e.][]{PRM08}. After segmentation, assigning discrete states to the segmented data is referred to as calling \citep{WielCall}.

Here, we will assume that the data has been properly segmented and discretized. While the measurement unit is clearly defined from the array platform used, there is no a priori biologically most relevant unit of interest in DNA copy number experiments. This contrasts the majority of mRNA experiments, for which the genes as coding regions for proteins are a natural relevant unit. By nature, the crucial DNA copy number events, aberrations, arise when a piece of DNA is either deleted or gained. Such a piece can be an entire chromosomal arm, but also just $1/10^6$th of an arm. Therefore, we aim to learn the relevant units from the data. Such units are particularly important for interpretation and power when testing association with clinical data. For simplifying terminology, we refer to the array CGH features as probes although this could also reflect clones or cDNAs. Then, our proposed two-step dimension reduction approach enables one to consider the data at two levels with decreasing resolution: regions and clusters.  The regions are the result of collapsing neighboring probes of which the discretized values are highly repetitious and redundant within (almost) all profiles \citep{WielRegions}, while the clusters contain contiguous similarly behaving regions.

After collapsing, our study proceeds in two directions. First, we develop a model-based clustering algorithm that considers spatial patterns. Second, we perform simultaneous region-wise and cluster-wise multiple testing. The idea of multiple testing based on clustered results is not new. \cite{benjamini07jasa} adapt the False Discovery Rate (FDR) to allow for cluster-level multiple testing in the analysis of functional Magnetic Resonance Imaging (fMRI) data. Moreover, comparing to voxel-wise (feature level) multiple testing, they asserted that their method improved SNR (signal to noise ratio) and gained statistical power.
Even though both array CGH data and fMRI data have strong spatial correlation patterns, we cannot apply the methodology of \cite{benjamini07jasa} directly. Using fMRI it is often feasible to obtain the clusters of voxels from independent, preparatory scans.  In array CGH data, however, it is hardly possible to obtain such information a priori. One solution is splitting the set of samples in two parts, where one is used for clustering and the other for testing, which guarantees independence between clustering and testing. However, in that case, we should accept significant power loss due to splitting. \cite{Pacifico2004} offer an alternative solution, also in the context of imaging data and FDR: clustering on the basis of $p$-values. Their theory is based on smooth Gaussian random fields, which is not a realistic model for array CGH data. Moreover, this would render meaningful clusters from a testing perspective only.

We show that, in the context of permutation testing and family-wise error rate (FWER), it is possible to apply both clustering and testing to the same data set if the cluster result is permutation invariant.
We consider selection of a suitable test statistic on both levels (regions and clusters) and use an hierarchical testing procedure to control the FWER. The entire clustering plus testing procedure is illustrated on two array CGH cancer data sets.

\subsection*{High-dimensionality}
\label{sec:high_dim}
High-dimensionality is a common theme for the analysis of data produced by high-throughput genomic technologies. It denotes the situation that the number of features is much larger than the number of samples. Due to the large dimension, traditional multiple testing procedures such as Bonferroni yield very conservative results.

While the term `high-dimensionality' is used neutrally in terms of the number of features, some authors note that technological developments in data generating processes can allow a more specific approach for the data. \cite{benjamini07jasa} suggest that the assumption of infinitely many elements in a cluster is not unrealistic for fMRI data in an environment of improved technical resolution. A similar development has occurred for our type of genomic data. Here, the resolution increased from chromosome arms ($\sim 40$) to Bacterial Artificial Clones ($\sim 3000$), to oligos ($< 400,000$), to next generation arrays ($> 1,000,000$) \citep{park08cancerinvest}.
The increasing number of features does not necessarily imply an increasing number of `distinct' relevant units, because these new features may appear as repetitions of similar copies on a coarser scale.

We briefly illustrate the above idea. Consider the extreme case in which a certain region of the genome spanning $r$ probes never contains a genomic breakpoint for the type of samples under study. Doubling the resolution in this genomic region to $2r$ will only lead to measuring more of the same (due the absence of a breakpoint) and hence for both resolutions the relevant unit can be collapsed to one data point per sample for this genomic region.

Discretized array CGH data usually appears as large numbers of distinct blocks which consist of probe vectors with
(almost) the same discretized status (across samples) for as many as hundreds or thousands of probes. So, regions in the data correspond to distinct blocks. Those regions may be considered as biologically relevant units for the type of samples under study.

Collapsing repetitious probes is useful, since it reduces unnecessary dimension of data and enhances computational convenience. Conventional clustering methods using correlational association among probes seem not appropriate in this stage, because these may yield mingled clusters: repetitiously copied probes mixed with similarly behaving probes. Hence this may weaken the performance and interpretation of the clustering methods usually performed in the next stage. From this point of view, it is desirable to separate two procedures: collapsing and clustering. The first step handles the technical resolution and is used for dimension reduction. The latter step is applied for combining similar behaving probe regions so that one may obtain hints of collective behaviors within certain genomic neighborhoods.

We do not suggest a new algorithm for the collapsing process, but instead use the one proposed by \cite{WielRegions}. Here, the relative amount of information lost by collapsing can be controlled; we use 0.5\% as an upper bound. Note that this algorithm maintains the high resolution where desired: small genomic segments that differ consequently from their neighbors are kept. In construction of the clustering models, we incorporate the collapsing information into our model in two ways: either via base-pair distance between two regions or via the number of collapsed probes between two regions.

% \subsection{Discrete multiple testing}
% \label{sec:discrete}
% 
% The goal of region-wise multiple testing in array CGH data is to find significantly expressed regions between two groups, a test group and a reference group. We consider a region-wise test of independence between groups and states. Data for each region is tabulated by a $2 \times 3$ contingency table. We perform nominal independence test for association between 2 groups and 3 states.
% 
% In principle, discrete data produces discrete test statistics and discrete $p$-value distribution. Actual multiple testing control level for discrete data are usually conservative, because individual achievable significance level is strictly smaller than nominal individual significance level so that overall control level becomes smaller. Also, depending on null distributions, $p$-value of each region-wise test may differ.
% 
% For different null assumptions, different implementations are suggested. Once we assume fixed row and column margins, Fisher's exact test is most desirable. If we assume fixed row margin alone, unconditional tests can be used but we still need to specify remaining parameters (nuisance parameters) which are not used in the testing. Or we may consider a permutational null distribution preserving correlation structures among regions. Since sample size of each group is small, we exclude the case of using asymptotic null distribution. For the detail, we explain those assumptions and implementations in the Methods section.

\section*{Methods}
\subsection*{Mathematical setting}
In this paper, the overall statistical model corresponds to $n$ i.i.d. copies of a pair of random variables denoted by $(X^1,Y^1)$, $...$, $(X^n,Y^n)$, where, for each individual $i=1,...,n$,  $X^i \in \{0,-1,1\}^m$ is the state vector along all the (ordered) regions and $Y^i \in \{0,1\}$ is a label determining which group the individual belongs to.
In this model $X=(X^1,...,X^n)$ is a sample of i.i.d. variables, in which the distribution of $X^i$ can be seen as a mixture between the two groups $Y^i=0$ and $Y^i=1$. In particular, note that the distribution of the $X^i$'s does not involve the relationship between the $Y^i$'s and the $X^i$'s.

In a nutshell, our approach firstly clusters the rows of $X$ that are similar from a model-based point of view and secondly detects the regions and clusters that are significantly associated to the label vector
$Y=(Y^1,...,Y^n)$:
\begin{itemize}
\item The clustering phase uses the variables $X$, by assuming a specific parametric model for the distribution of the $X^i$'s .  Since $X^1,...,X^n$ are i.i.d., the clustering result $\widehat{\mathcal{A}}(X)$ is invariant under permutations of the columns of $X$.
\item The testing phase tests the independence between
  regions
  $X_j^{\sbullet}=(X_j^i)_{1\leq i \leq n}$ and $Y$
  and between clusters $(X_j^{\sbullet})_{j \in A}$ and $Y$,
  conditional on the clustering  $\widehat{\mathcal{A}}(X)$. The (conditional) $p$-values are computed by performing permutations of the observed labels, which is valid because $\widehat{\mathcal{A}}(X)$ is permutation invariant. Then, these $p$-values are integrated in an hierarchical multiple testing procedure controlling the family-wise error rate both on the region and cluster level.
\end{itemize}

\subsection*{Clustering model}

\begin{figure}[htbp]
  \begin{center}
    \includegraphics[width=.45\textwidth]{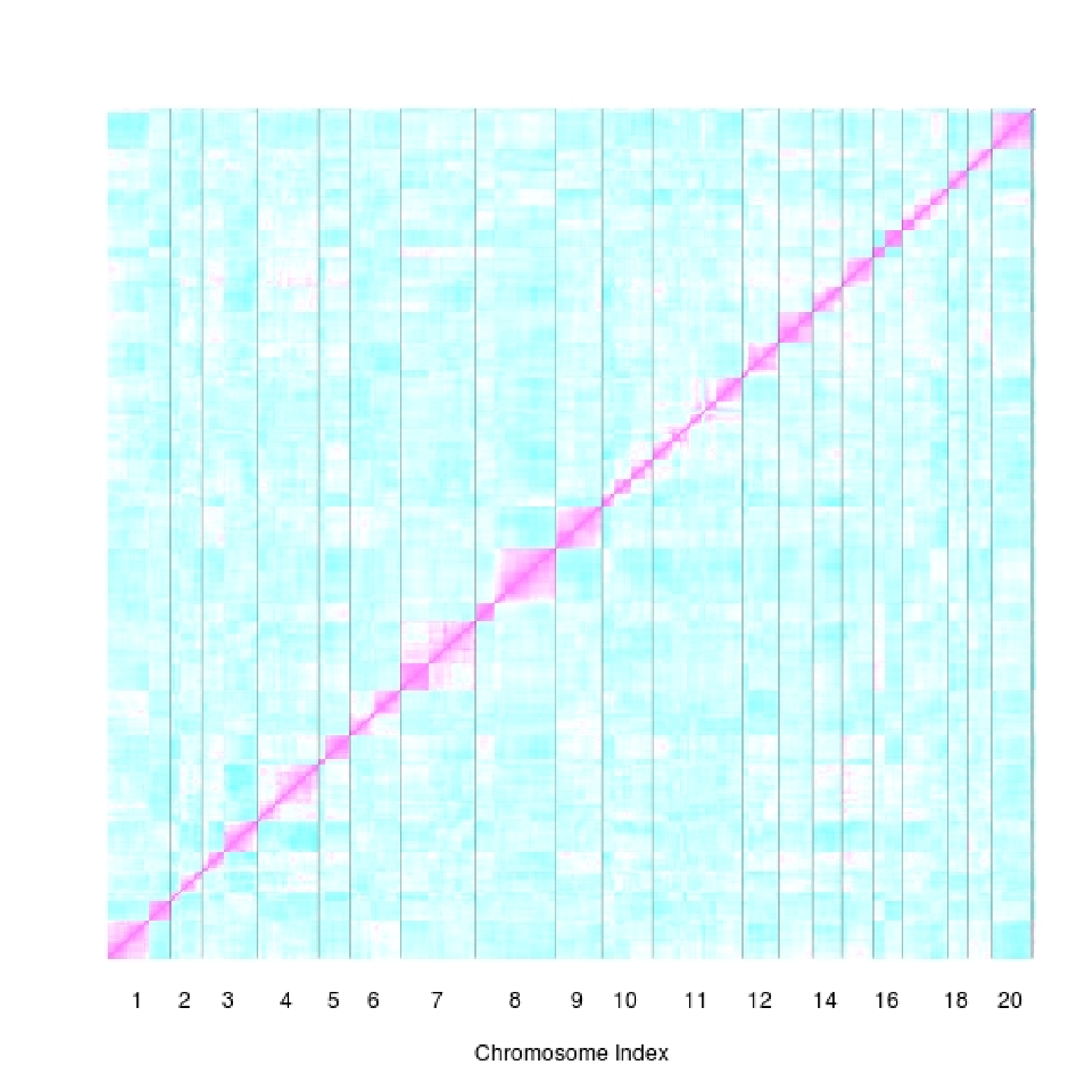}
    \includegraphics[width=.45\textwidth]{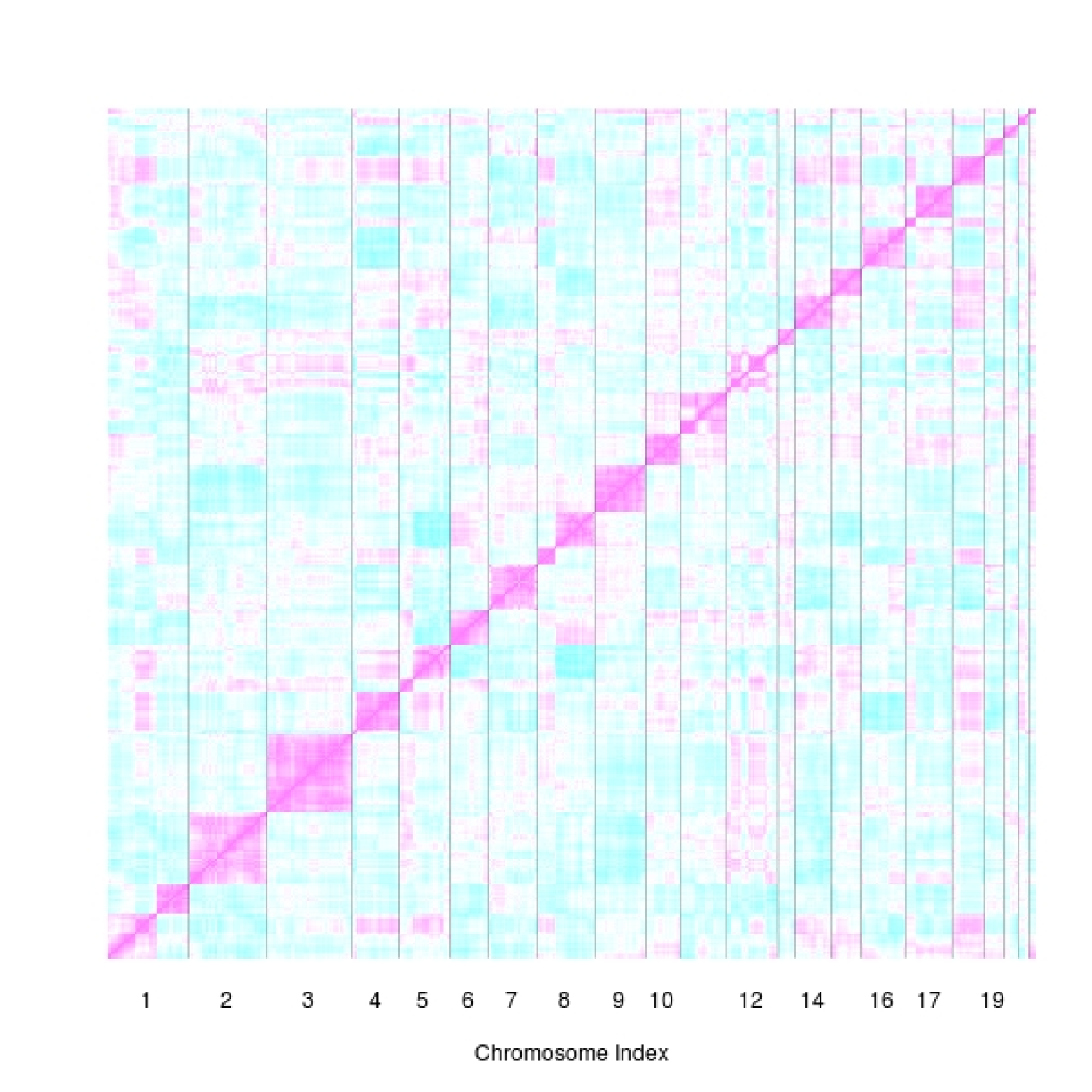}

    \caption{Two correlation (Kendall's $\tau$) heatmaps for \cite{Chin2006} (left) and \cite{douglas04cancerres} (right) data sets. Data sets contain $383$ and $436$ regions, respectively. Regions are plotted according to chromosomal position. Colors represent correlations from -1 (cyan) to 1 (pink).}
    \label{fig:entropy}
  \end{center}
\end{figure}

Our motivation for spatial clustering comes from Figure~\ref{fig:entropy}. In Figure~\ref{fig:entropy}, we illustrate Kendall correlations for pairs of all regions for two discretized array CGH data sets. Strong correlations are concentrated in the diagonal parts, so spatial dependency patterns are notably block-wise structures. Such correlations are caused by a high likelihood of the same aberration to span multiple consecutive regions (as opposed to off-diagonal correlations between regions on different chromosomes, which necessarily represent different aberrations). We incorporate the spatial correlation in our modeling approach. Log-linear models are a natural candidate for describing discrete data. Hence, we apply such models for our purpose, while adapting these to take specific data characteristics such as spatial correlations, but also physical base pair distance between regions, into account. Moreover, we consider practical concerns for efficient optimization of the clustering.

Let $m$ be the number of regions and $n$ be the number of i.i.d. samples under consideration. Moreover, let each region be indexed by its spatial location on the DNA, that is, the $(j-1)$-th and $(j+1)$-th regions are the neighbors of the $j$-th region. The goal is to identify clusters so that regions in one cluster behave similarly.

We first consider the quadratic exponential model \cite[]{cox94biometrika} which approximates log-linear models for binary data up to 2nd order interactions. In our case we consider trinary valued variables. Because the samples are initially assumed i.i.d., we suppress the sample index when describing the model, so $X_j=x_j$ refers to the observed state of the $j$th region. Hence, for observations $x_A = (x_j)_{j\in A},$ with $x_j \in \{0, -1, 1\}$ and $A \subset \{1,\ldots,m\}$, the marginal distribution of the corresponding random variable $X_A$, $p_A$, is modeled as
\begin{equation*}
  \log p_A(x_A) = \alpha_A + \sum_{j \in A}\beta_{A,j} x_j + \sum_{j < k,\: j,k \in A} \gamma_{A, jk} f(x_j,x_k),
\end{equation*}
where $f(x_j,x_k)= -1$, if $x_j \neq x_k$ and $f(x_j,x_k)= x_j x_k$, otherwise.
Note that $\alpha_A$ is not a free parameter, because it is determined by the `summing-to-one' condition.
The discrete values `$0$', `$-1$' and `$1$' denote normal ($=$ 2 copies), loss ($<$ 2 copies) and gain ($>$ 2 copies), respectively. We do not consider amplification (high copy number gain) and double deletion (0 copies) as separate states because of computational complexity of the clustering algorithm, but alternative solutions to accommodate such different states are discussed in the Discussion.

Since our purpose is not to specify every 2nd order interaction $\gamma_{A, jk}$, but to bind features which are in the same cluster, it may be better to simplify the above model by assigning one average 2nd order interaction per cluster. Since in actual array CGH data we observe physical base-pair distance between two regions, we use this information in the model. We assume the dependence of two regions becomes weaker for more distant pairs of regions (similar to p.430 of \cite{cressie93wiley}). Hence the interaction parameter $\gamma_{A,jk}$ is represented as $\gamma_A \tilde{d}_{jk}$ and
\begin{eqnarray*}
  \tilde{d}_{jk} = \left\{
    \begin{array}{cl}
      d_{jk}^{-q}/\max\{d_{jk}^{-q}\} & \text{if $j$ and  $k$ are in the same chromosome,} \\
      0 & \text{otherwise.}
    \end{array}
  \right.
\end{eqnarray*}
where $d_{jk}$ is the base-pair distance between two regions $j$ and $k$ and the maximum is computed over pairs in the same chromosome. Alternatively, the distance function can be used to incorporate the collapsing information; for example, the distance between two regions $j$ and $k$ can be the actual number of probes (measurement unit) between them. Here, $q$ is the decreasing rate of spatial dependence. Common choices for $q$ are $0, 1, 2$ \citep{cressie93wiley}. Let $\theta_\l = (\alpha_{A_\l}, \beta_{A_\l,1}, \ldots, \beta_{A_\l,m_\l}, \gamma_{A_\l}),$ where $A_\l$ is a set of contiguous regions and $m_\l$ is the number of regions in $A_\l$. Then, our full model for a partition $\mathcal{A}=\{A_\l\}_{\l=1}^{K_{\mathcal{A}}}$ of $\{1,\ldots,m\}$, where $K_{\mathcal{A}}$ denotes the number of clusters, is
\begin{equation}
  \label{eq:qem}
  p_{\mathcal{A}}(x) = \prod_{\l=1}^{K_{\mathcal{A}}} p_{A_\l}(x_{A_\l}; \theta_\l),
\end{equation}
with
\begin{equation}\label{eq:qemb}
  \log p_{A_\l}(x_{A_\l}; \theta_\l)= \alpha_{A_\l}+\sum_{j \in A_\l} \beta_{A_\l,j}x_j + \gamma_{A_\l} \sum_{j<k, j,k \in A_\l} \tilde{d}_{jk} f(x_j, x_k).
\end{equation}
From \eqref{eq:qem}, we know that $X_{A_\l}$ and $X_{A_{\l'}}$ are assumed independent for different $\l$ and $\l'$. 
Denote the value of $x_{A_\l}$ for sample $i =1, \ldots, n$ by $x^i_{A_\l}$.
Then, we search a partition and parameter setting that optimizes the following log-likelihood:
\begin{equation}
  \label{eq:optim}
  (\hat{\mathcal{A}},\hat{\theta}) = \argmax_{\mathcal{A} \in \mathcal{C},\theta} \sum_{i=1}^{n}\sum_{A_\l \in \mathcal{A}} \log p_{A_\l}(x^i_{A_\l}; \theta_\l)
\end{equation}
where $\mathcal{C}$ is the set of contiguous partitions and $\theta = (\theta_\l)_{\l=1}^{K_{\mathcal{A}}}$.
Note that the model (\ref{eq:qem}) contains an intrinsic trade-off for dividing a cluster into two subclusters:
potentially higher likelihood due to an extra parameter $\gamma$ ($\alpha$ is not a free parameter) with respect to one cluster versus potentially lower likelihood due to the forced independency for two regions in a different subcluster. Which effect prevails depends on how similar the regions between two subclusters are. As a consequence, the number of clusters $K_{\mathcal{A}}$ is not defined a priori, but instead computed through the optimization step.

Generally, searching the space $\mathcal{C}$ for large $m$ is computationally intractable. Therefore, we restrict the maximum size of a cluster. The components of the product in the right hand side of \eqref{eq:optim} are computed in two steps; first, we estimate the parameters of each cluster smaller than the maximum size. From these parameters, the likelihood of each candidate contiguous partition can be determined. Second, given the likelihoods, we search the partition space and, due to the specific partition structure \citep{cormen00mit}, we achieve the global optimum for the equation \eqref{eq:optim} by applying Dijkstra's shortest path algorithm. Note that, from a biological perspective, a cluster should not cover more than one chromosome, so we apply the algorithm per chromosome separately.

One could question the usefulness of this cluster algorithm when the results would be sensitive to similar region-wise differences between, say, two clinical groups. However, observe from (\ref{eq:qemb}) that the likelihood for a given partition $A_\l$ is to a large extent determined by within-sample similarity. This means that, as opposed to a $p$-value type clustering method, our method is unlikely to cluster two independent consecutive regions that coincidentally have a similar group difference, because the independence implies that these two differences are likely to be caused by (partly) different samples. Moreover, in terms of causality, the perception is that the DNA copy number data (which are early onset markers) potentially affects the group labels, rather than the other way around.
Nevertheless, in rare cases, two independent consecutive regions may have realizations  that coincide for many samples, in particular when the marginal distributions within groups would (almost) degenerate (e.g. all 0's in one group and all 1's in the other). This may cause coincidental clustering. Therefore, we investigate how likely such coincidental clustering is for a given data set using a shuffling argument.

Denote the event that a random pair of consecutive regions clusters by $E$, and the event that these two regions are dependent due to their genomic proximity by $D$. Its complement, lack of genomic dependency, is denoted by $D'$. If our cluster method performs properly $P(D|E)$ should be very high. Applying Bayes' rule, we have
\begin{equation}\label{Bayesrule}
  P(D|E) =
  \frac{P(E|D)P(D)}{P(E)} = \frac{P(E|D)P(D)}{P(E|D)P(D) + P(E|D')P(D')}.
\end{equation}
From the data we easily estimate $P(E)$ by counting consecutive pairs of regions that cluster.  Then, if $P(E|D')$ is very small and $P(E)$ is fairly large, we necessarily have: $P(E|D)P(D) \gg P(E|D')P(D')$, which implies $P(D|E) \approx 1$. To show that $P(E|D')$ is small, we break the genomic dependency structure by shuffling regions such that two regions of the same chromosome are not allowed to be neighbors in the shuffled data set. Then, we apply our clustering method to the shuffled data set. If very few regions cluster, then, necessarily, $P(E|D')$ is small. The shuffling is repeated several times to account for the arbitrariness of the shuffling.

\subsection*{Clustering stability}
\label{sec:stability}
Since a clustering result depends on the data used, especially on the sample size, there exists uncertainty on the clustering result. We suggest to use the adjusted Rand index \citep{hubert85jc} to evaluate stability of the clustering results. If a small change of the data results in large change of the clustering result, then the clustering method is instable. Therefore, we check the stability using a cross validation framework.

Suppose regions are numbered from $1$ to $m$ and a clustering result is represented as a contiguous partition  of $\{1,\ldots,m\}$ as in the previous section. Let $\mathcal{A}=\{A_\l\}_{\l=1}^{K_{\mathcal{A}}}$ and $\mathcal{A}'=\{A'_{\l'}\}_{\l'=1}^{K_{\mathcal{A}'}}$ be two such partitions which we aim to compare and let $|A|$ be the cardinality of index subset $A$. Then the adjusted Rand index (ARI) for the two partitions is defined as
\begin{eqnarray*}
  \text{ARI}(\mathcal{A},\mathcal{A}') = \frac{r-E(R)}{\max(R)-E(R)},\quad r = \sum_{\l=1}^{K_{\mathcal{A}}} \sum_{\l'=1}^{K_{\mathcal{A}'}} \binom{|A_\l \cap A'_{\l'} |}{2},
\end{eqnarray*}
where $r$ is the observed value of random variable $R$. The expectation of $R$ is computed conditional on the fixed margins, $|A_\l|,|A_{\l'}|,\l=1,\ldots, K_{\mathcal{A}},\: \l'=1,\ldots,K_{\mathcal{A}'}$ and the maximum is computed as the maximum of $R$ regardless of the fixed margins. The adjusted Rand index takes numbers between $-1$ to $1$.

After obtaining $V$ clustering results from $V$-fold cross validation, we compute the average adjusted Rand index, aARI, which serves as our stability measure:
\begin{equation}\label{meanARI}
  \text{aARI} = \frac{1}{V}\sum_{v=1}^V \text{ARI}(\mathcal{A},\mathcal{A}^{(v)})
\end{equation}
where $\mathcal{A}$ is the original clustering result from the full data and $\mathcal{A}^{(v)}$ is the $v$-th cross-validated clustering result.

\subsection*{Hierarchical multiple testing for clusters and regions}
\label{sec:stat}
Once we have the clusters $\mathcal{A}=\{A_\l\}_{\l=1}^{K_{\mathcal{A}}}$, our focus turns to testing association of the region and cluster-wise data with clinical information.
Let us first shortly consider association tests between groups (reference and test) and states (loss, normal and gain) for a $2 \times 3$ contingency table $n = (n_{11}, n_{12}, n_{13}, n_{21}, n_{22}, n_{23})$ as Table~\ref{tab:cont}.
\begin{table}[htbp]
  \centering
  \begin{tabular}{c|c|c|c|c}
    group $\backslash$ state & loss & normal & gain & \\
    \hline
    reference & $n_{11}$ & $n_{12}$ & $n_{13}$ & $n_{1+}$\\
    \hline
    test & $n_{21}$ & $n_{22}$ & $n_{23}$ & $n_{2+}$ \\
    \hline
    & $n_{+1}$ & $n_{+2}$ & $n_{+3}$ & $n_{++}$\\
  \end{tabular}
  \caption{Contingency table representation of copy number aberrations between two groups.}
  \label{tab:cont}
\end{table}
Let $\pi_{gs}$ be the population proportion for group $g$ and state $s$. Then, for each region $j$  we test \begin{equation}\label{H0reg}
  H_{0j}: \mbox{ for any }g,s,\: \pi_{gs,j} = \pi_{g,j} \pi_{s,j}.
\end{equation}
For a cluster $A_\l$, we simultaneously test
\begin{equation}\label{H0cl}
  H_{0}^\l: \bigcap_{j\in A_\l} H_{0j}.
\end{equation}

Besides the usual scala of test statistics for a particular testing problem, two classes of approaches may be distinguished: unconditional and conditional tests. In the first case, usually only the sample sizes (row margins) are fixed while column margins are variable. In the second case, both margins are fixed. Permutation tests fall in the latter category, since these only permute the labels (reference or test) for all samples. Permutation tests are useful to approximate (summaries of) multivariate distributions, which is exactly what we need for testing association on the level of clusters. Therefore, we use permutation tests in combination with the popular Pearson $X^2$ statistic for determining the significance of regions and clusters.

A test statistic for testing (\ref{H0cl}) is $M_\l = \min_{j\in A_\l} p_{j}$, where $p_j$ is the $p$-value obtained for region $j$ in cluster $A_\l$. Using the $p$-values to define the cluster test statistic standardizes the regions within a cluster. Before further motivating the use of $M_\l$, we first introduce the hierarchical testing approach.

The hierarchical multiple testing procedure we propose to use for the cluster-region data is based on that by \cite{Meinshausen2008}. Such a procedure allows to test both clusters and regions within one multiple testing framework. The procedure controls Family-Wise Error Rate (FWER) for hierarchical hypotheses. Why FWER and not the more popular False Discovery Rate (FDR)? Firstly, on the level of regions (usually a few hundreds) FDR may not reflect what it should, because a large `cluster' of highly correlated regions could have a disproportional large share in the number of discoveries, which highly impacts the estimated FDR for other regions. This is a consequence of the fact that FDR does not provide control on subsets \citep{Finner2001}. Secondly, our constructed clusters are more independent than regions, but their number is much lower (often below 100). Then, the power enhancement of an FDR procedure w.r.t. FWER is usually quite subtle, which may not outbalance the stronger conclusion one is allowed to draw with FWER control.

The procedure by \cite{Meinshausen2008} assumes an a priori known clustering. Such a setting can be relevant for these data as well, but often it is preferable to use the same data for both clustering and testing (see Discussion).
In a clustering plus permutation-testing framework, control of the FWER conditional to the data-based clustering
is feasible when the clustering is permutation invariant: when testing $H_{0}^{\l}$ using permutations, we permute the clinical responses (group labels
$Y$ in our case), while keeping the regions $X_j^{\sbullet}$ fixed for $j \in A_{\l}$.  If the result of the cluster algorithm is column-wise permutation invariant, the clusters may, in the testing phase, be assumed to be known when the null-hypotheses are formulated conditionally to the clusters. More mathematical details on the validity of permutation tests in this setting are presented in Appendix II. A similar argument for combining permutation testing with clustering is given by \cite{Goeman2010}.

The cluster method introduced in this paper is permutation-invariant, because it does not use the group labels. We use the sequential hierarchical testing procedure proposed by \cite{Goeman2010}, with critical values that depend on the rejection set $\mathcal{R}$, which contains both the rejected clusters and regions. This procedure is a slightly more powerful alternative to the one by \cite{Meinshausen2008}. It applies the so-called inheritance principle in combination with the Shaffer (\citeyear{Shaffer86}) improvement in an hierarchical testing context. The inheritance principle is a variation of the fall-back principle \citep{Wiens2005} that allows to test an hypothesis less strictly (applying a larger $p$-value threshold) when neighboring hypotheses are rejected. Likewise, the Shaffer improvement allows use of a larger $p$-value threshold for regions in a cluster, because it uses the connection between the cluster and region hypotheses: if $H_{0}^\l$ is false, at least one $H_{0j}$ should be false too for $j\in A_\l$ (see Appendix II). We emphasize that this procedure guarantees strong control of FWER \citep{Goeman2010}. It relies only on the individual permutation $p$-values of clusters and regions. It is a Holm-type procedure and hence subset pivotality \citep{DudoitShaffer} is not required.
% Since the joint distribution of $p$-values may be different under the complete null and under the partial null,  may not hold an

For a given rejection set $\mathcal{R}$ (both for clusters in $\mathcal{A}$ and regions in $\{1,...,m\}$), critical values for clusters are defined by $\alpha_{\mathcal{R}} = \alpha/(K_{\mathcal{A}} - D_{\mathcal{R}}),$ where $D_{\mathcal{R}}$ equals the number of clusters in $\mathcal{R}$ for which all regions are members of $\mathcal{R}$ as well. Note that we opt, as opposed to \cite{Meinshausen2008}, to weigh clusters equally, because for our application small clusters may be as relevant as large ones. This does not affect control of FWER (see Appendix III for a proof).  Critical values for regions in cluster $A_\l$ are denoted by $\alpha_{\mathcal{R},\l} = \alpha_{\mathcal{R}}/(|A_\l|-\max(1,D_{\mathcal{R},\l})),$ where $D_{\mathcal{R},\l}$ equals the number of regions in cluster $A_\l$ that are members of $\mathcal{R}$.

Then, the hierarchical testing procedure, which is initiated by $\mathcal{R} = \emptyset$, %varnothing$,
is as follows.

\begin{enumerate}
\item Reject $H_{0}^{\l}$ for cluster $A_\l$ if $P_{0}(M_\l \leq m_\l) \leq \alpha_{\mathcal{R}},$ where $m_\l$ is the realization of $M_\l$.
\item If $H_{0}^{\l}$ is rejected, reject $H_{0j}$ for region $j$ in cluster $A_\l$ if
  $p_j \leq \alpha_{\mathcal{R},\l}$. % where $X_r$ is the $X^2$ test statistic and $x_k$ is its realization.
\item Update the rejection set $\mathcal{R}$ and critical values $\alpha_{\mathcal{R},\l}$.
\item Repeat steps 2 and 3 for the non-rejected regions until no more regions are rejected.
\item Update cluster critical values $\alpha_{\mathcal{R}}$.
\item Repeat steps 1 to 5 for the non-rejected clusters and regions therein.
\item Stop when no hypothesis is rejected anymore.
\end{enumerate}

One could argue that for testing clusters $M_\l= \min_{j\in A_\l} p_{j}$ may have less power than a statistic that focuses more on `average behavior' (such as a median $p$-value or a sum of standardized region-wise test statistics). This is true when small effects add up to one larger cluster-wise effect, which is quite common in mRNA gene expression studies. However, we believe the following scenarios to be more relevant for the clustered array CGH data: 1. a cluster is homogenous (large $\gamma$), and hence high positive dependencies between regions within the cluster are present;
2. a cluster is heterogenous, and for only a few regions the association exists. It is clear that in the first scenario little power is lost when using $\min p$ with respect to f.e. $\text{median}\,p$, while in the second scenario $\text{median}\,p$ has less power than $\min p$. Using the hierarchical testing approach, the latter could, after rejection of $H_{0}^\l$, still identify significant regions in the heterogenous cluster.

\section*{Results}
We analyze two data sets from \cite{Chin2006} (Data1) and from \cite{douglas04cancerres} (Data2). Both data sets have been discretized to ternary values, $0$, $-1$ and $1$ \citep{WielCall}. After collapsing, Data1 contains 383 regions in rows and 96 and 49 samples in columns, which correspond to ER-positive and ER-negative breast cancer samples, respectively. Data2 contains 436 regions in rows and 7 and 30 samples in columns involved in colorectal cancer, representing two group states: microsatellite instable and chromosomal instable, respectively.

We first concentrate on the clustering results. For the clustering algorithm, the maximum number of regions per cluster is constrained to $9$, which we generally found to be sufficiently large. Parameter $q$, the decreasing rate of spatial dependence is set to $1$. Finally, $\gamma$ is re-scaled by $\tilde{\gamma} = (e^{\gamma}-1)/(e^{\gamma}+1)$ so that it lies between $-1$ to $1$ as the nominal correlation coefficient (equation (6.5.10) in p.430 of \cite{cressie93wiley}).

\begin{figure}[]
  \begin{center}
    \includegraphics[height=.5\textheight]{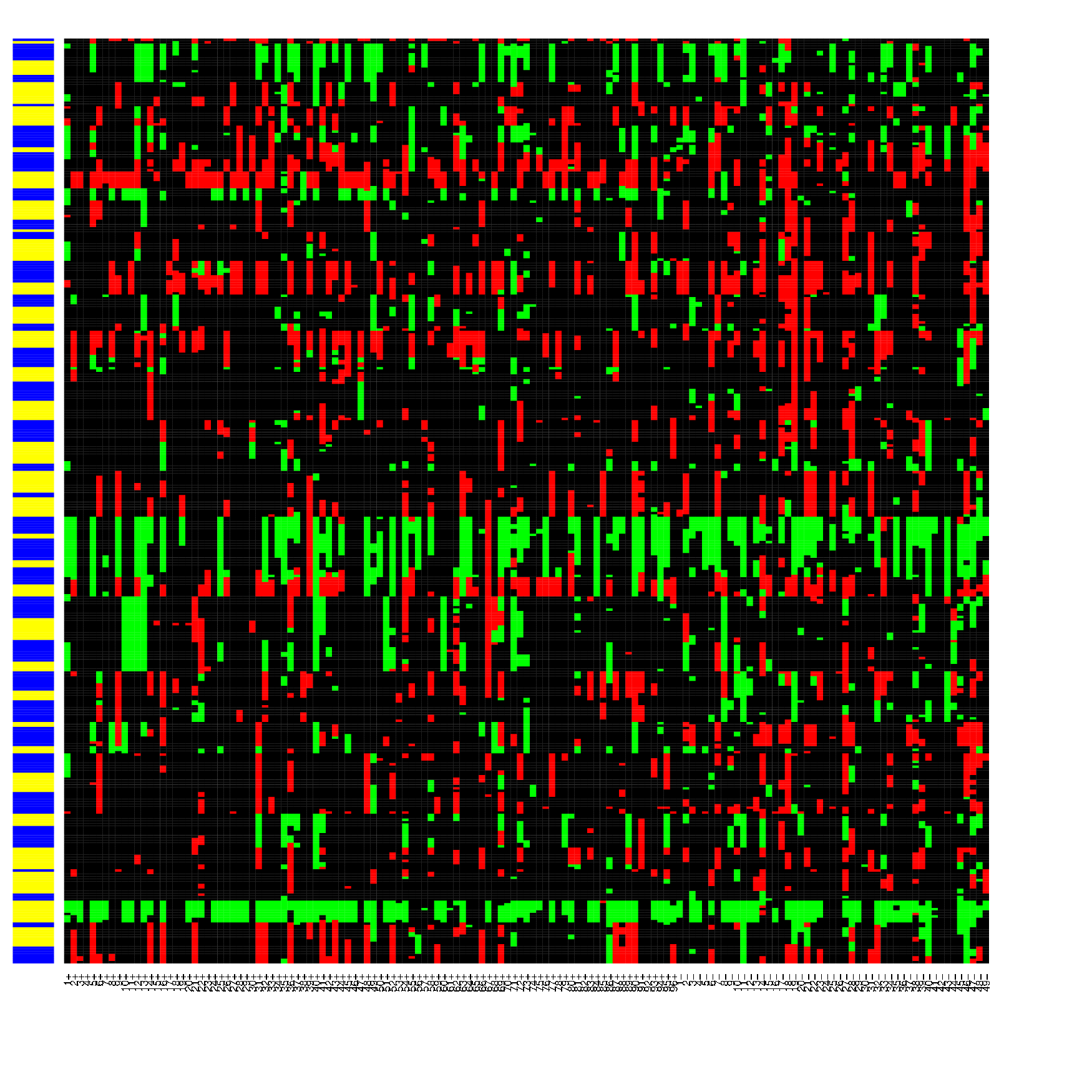}
    \caption{Clustering results for Data1. Losses are plotted in red, normals in black, gains in green. Clusters are order according to chromosomal position from bottom to top and depicted alternately in yellow and blue. Sample labels are plotted on the bottom axis, ``+'' indicates an ER-positive sample, ``-'' an ER-negative one.}
    \label{fig:data1}
  \end{center}
\end{figure}

\begin{figure}[]
  \begin{center}
    \includegraphics[height=.35\textheight]{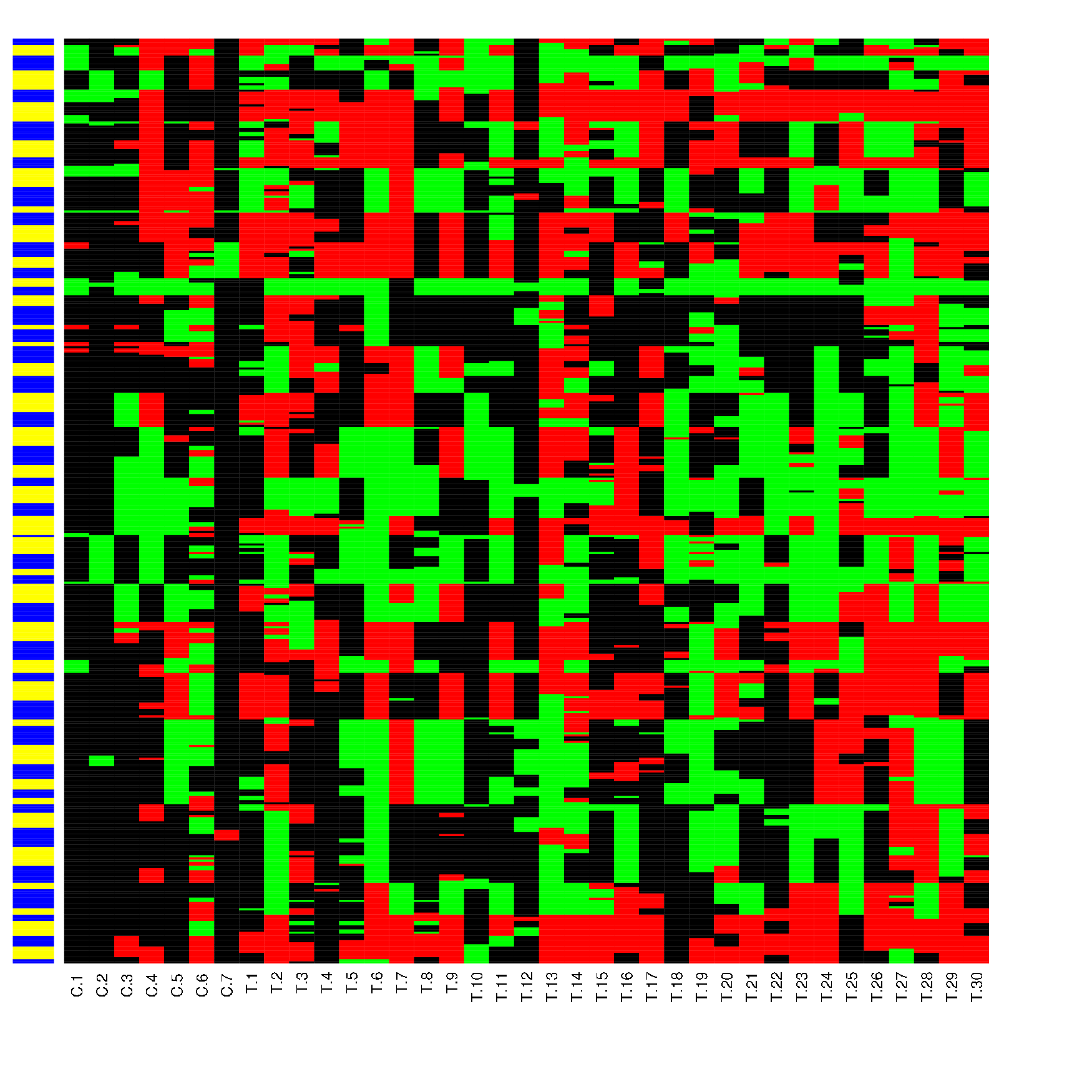}    
    \caption{Clustering results for Data2. See caption Figure \ref{fig:data1}.}
    \label{fig:data2}
  \end{center}
\end{figure}

Figures \ref{fig:data1} and \ref{fig:data2} illustrate the clustering results for Data1 and Data2, respectively. The number of clusters is 63 and 69, respectively. The 10-fold cross-validation stability score, the adjusted average Rand index (\ref{meanARI}), is high for both data sets: $0.969$ and $0.963$, respectively. These indicate that the cluster results do not strongly depend on the in- or exclusion of 10\% of the samples.
% The stability scores from bootstrapping the data 300 times are of course lower: $0.937$ for Data1 and $0.870$ for Data2, respectively. The latter is somewhat lower due to the smaller size of the second data set.

For both data sets we also investigate the probability on coincidental clustering. More precisely, we show that for both data sets the probability that two (consecutive) regions in a cluster are really dependent is high. Following the argument outlined in the Methods section (see equation (\ref{Bayesrule})) we first need to show that the probability that two consecutive regions cluster, $P(E)$, is fairly large. Given the relatively small number of clusters in both data sets, this is the case. Next, we need to show that probability that two independent regions cluster, $P(E|D')$, is low by considering shuffled data sets. Indeed, using 25 shuffled data sets, we observe that for Data 1 (383 regions), on average 382.2 (range: 381-383) clusters are formed, while for Data 2 (436 regions) on average 434.5 (range: 432-436) clusters are formed. Hence, we are confident that clusters in the original data sets are almost always created because of genomic spatial dependency.

% Note that the $\tilde{\gamma}$ values are all positive, since, although the data is encoded by trinary values actual data per region shows mostly binary features, such as $0/1$ or $0/-1$.

\begin{figure}
  \begin{center}
    \includegraphics[height=.35\textheight]{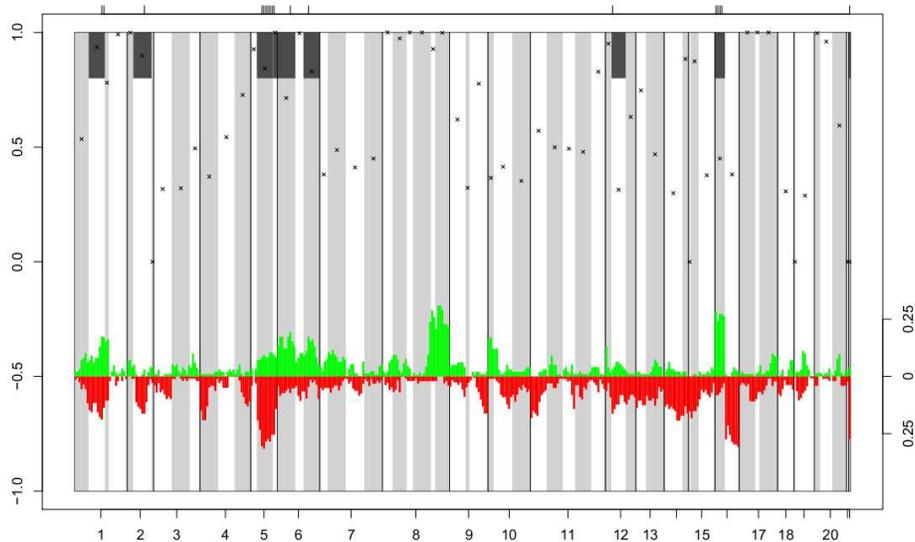}
    \caption{Clustering plus testing, Data1. Chromosomes on the bottom axis, separated by solid lines. Alternating light-grey and white bars demarcate clusters.
      Crosses show the association coefficients $\tilde{\gamma}$ (scale on left axis).
      Dark-grey bars: significant clusters, tick marks at the top axis: significant regions ($\alpha=0.1$). 
      Bottom plot: absolute difference between proportions of gains in the two groups (green) and between proportions of losses (red) (scale on right axis).
    }
    \label{fig2:data1}
  \end{center}
\end{figure}

\begin{figure}
  \begin{center}
    \includegraphics[height=.35\textheight]{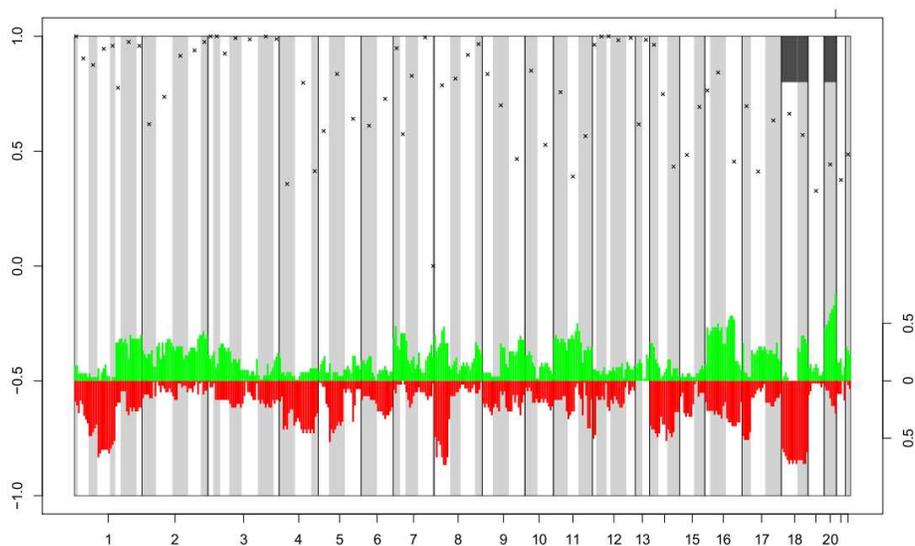}
    \caption{Clustering plus testing, Data2. See caption Figure \ref{fig2:data1}.}
    \label{fig2:data2}
  \end{center}
\end{figure}

For testing association of the class labels with the group labels, 20.000 permutations of the group labels are performed. The above hierarchical testing procedure was applied to the clusters and regions. Tables \ref{tab:data1} and \ref{tab:data2} contain the results of two significant clusters for both data sets, whereas Figures \ref{fig2:data1} and \ref{fig2:data2} combine the overall collapsing, clustering and testing results. Note that for both data sets the two groups are rather unbalanced. Hence, when a peak in the differential proportion of either gains or losses is caused by a large proportion of that aberration in the smaller group, these tend to be less significant (f.e. second cluster on chromosome 16, Figure \ref{fig2:data1}).

To elucidate the potential benefit of the clusters in a testing setting we compared the results of the hierarchical testing procedure with a simple Holm step-down procedure applied to the regions alone. Two FWER cut-offs were considered: $\alpha=0.05$ and $\alpha=0.1$. Our findings can be summarized as follows. For both data sets and both cut-offs the Holm procedure only identifies regions that are part of a cluster identified as significant by the hierarchical procedure. For Data1, $\alpha=0.05$, the hierarchical procedure identifies 8 clusters, two of which contain no regions that are identified by the Holm procedure. Holm identifies 13 regions, hierarchical two extra, 15.
For Data1, $\alpha=0.1$, the hierarchical procedure identifies 9 clusters, for which each contains at least one region that is identified by the Holm procedure. Both procedures identify the same 18 regions. For Data2, $\alpha=0.05$, the hierarchical procedure identifies 1 significant cluster of which none of the regions are identified by Holm. No significant regions are identified by both.
For Data2, $\alpha=0.1$, the hierarchical procedure identifies 3 significant clusters. Two clusters contain not a single significant region according to Holm. Both Holm and hierarchical identify (the same) one region. In summary, both procedures are comparable in terms of the number of identified regions, but the hierarchical procedure has a clear advantage: it is able to identify clusters of which none of the regions are identified by Holm.
\cite{Meinshausen2008} shows similar results using an hierarchical Bonferroni-type procedure for other data types.

We also compared the hierarchical procedure with a Holm step-down procedure on clusters. As expected, the number of detections using these procedures differs very little, because clusters are the highest level in the hierarchy and
the proportion of differential clusters is small.

% As an illustration, we show Table~\ref{tab:Data1} with $p$-values for regions in cluster 2 of Data1. Cluster 2 contains 7 significant regions but the association coefficient $\tilde{\gamma}$ is $0.9074$. First, for the individual region-wise testing result, 7 regions are detected as significant ann the result of all three distributional assumptions (unconditional, conditional, permutational) agrees. Next, for the clustering result, the association coefficient appears as high as $0.9074$ and this indicates the regions in cluster 2 are highly associated each other. Considering the spatial dependence pattern mentioned before and the high association coefficient of this cluster, we may need investigate the region 9 and 10 which are surrounded by significant regions further as potentially important regions.
% 
\begin{table}[htbp]
  \centering
  \caption{Comparison of $p$-values for two clusters in Data1. 
    Columns 2-4 and 7-9 denote the raw (unadjusted), Holm and hierarchical $p$-values for clusters and regions, respectively.
    Fifth column contains the association coefficient $\tilde{\gamma}$. 
  }
  \label{tab:data1}
  {\setlength{\tabcolsep}{4pt}
    \begin{tabular}{|c|c|c|c|c|c|c|c|c|}
      \hline
      Clust.	&	$p$ clust.	&	$p$ clust.	&	$p$ clust.	&	$\tilde{\gamma}$	&	Reg.	&	$p$ reg. &	$p$ reg. & $p$ reg.	\\
      & raw & Holm & Hier. & & & raw & Holm & Hier.\\\hline
      6	&	0.0006	&	0.0366	&	0.0402	&	0.4653	&	1	&	0.0058	&	1.0000	&	1.0000	\\
      6	&	0.0006	&	0.0366	&	0.0402	&	0.4653	&	2	&	0.0076	&	1.0000	&	1.0000	\\
      6	&	0.0006	&	0.0366	&	0.0402	&	0.4653	&	3	&	0.0023	&	0.8004	&	0.9246	\\
      6	&	0.0006	&	0.0366	&	0.0402	&	0.4653	&	4	&	0.0007	&	0.2541	&	0.3283	\\
      6	&	0.0006	&	0.0366	&	0.0402	&	0.4653	&	5	&	0.0002	&	0.0738	&	0.0938	\\
      6	&	0.0006	&	0.0366	&	0.0402	&	0.4653	&	6	&	0.0403	&	1.0000	&	1.0000	\\
      6	&	0.0006	&	0.0366	&	0.0402	&	0.4653	&	7	&	0.2804	&	1.0000	&	1.0000	\\
      6	&	0.0006	&	0.0366	&	0.0402	&	0.4653	&	8	&	0.1696	&	1.0000	&	1.0000	\\\hline

      16	&	0.0000	&	0.0000	&	0.0000	&	0.7999	&	1	&	0.0034	&	1.0000	&	0.2245	\\
      16	&	0.0000	&	0.0000	&	0.0000	&	0.7999	&	2	&	0.0009	&	0.3060	&	0.1139	\\
      16	&	0.0000	&	0.0000	&	0.0000	&	0.7999	&	3	&	0.0000	&	0.0000	&	0.0000	\\
      16	&	0.0000	&	0.0000	&	0.0000	&	0.7999	&	4	&	0.0000	&	0.0000	&	0.0000	\\
      16	&	0.0000	&	0.0000	&	0.0000	&	0.7999	&	5	&	0.0001	&	0.0374	&	0.0268	\\
      16	&	0.0000	&	0.0000	&	0.0000	&	0.7999	&	6	&	0.0001	&	0.0374	&	0.0268	\\
      16	&	0.0000	&	0.0000	&	0.0000	&	0.7999	&	7	&	0.0001	&	0.0189	&	0.0201	\\
      16	&	0.0000	&	0.0000	&	0.0000	&	0.7999	&	8	&	0.0001	&	0.0189	&	0.0201	\\
      \hline
    \end{tabular}}
\end{table}

\begin{table}[htbp]
  \centering
  \caption{Comparison of $p$-values for two clusters in Data2. See caption Table \ref{tab:data1}.
    % First 3 columns denote the raw, Holm and hierarchical $p$-value for the clusters. Fourth column contains the association coefficient $\gamma$. Last 3 columns denote the raw, Holm and hierarchical $p$-value for regions within the clusters.
  }
  \label{tab:data2}
  {\setlength{\tabcolsep}{4pt}
    \begin{tabular}{|c|c|c|c|c|c|c|c|c|}
      \hline
      Clust.	&	$p$ clust.	&	$p$ clust.	&	$p$ clust.	&	$\tilde{\gamma}$	&	Reg.	&	$p$ reg. &	$p$ reg. & $p$ reg.	\\
      & raw & Holm & Hier. & & & raw & Holm & Hier.\\\hline
      66	&	0.0013	&	0.0863	&	0.0875	&	0.7076	&	1	&	0.0006	&	0.2371	&	0.1925	\\
      66	&	0.0013	&	0.0863	&	0.0875	&	0.7076	&	2	&	0.0065	&	1.0000	&	0.9030	\\
      66	&	0.0013	&	0.0863	&	0.0875	&	0.7076	&	3	&	0.0017	&	0.7174	&	0.3570	\\
      66	&	0.0013	&	0.0863	&	0.0875	&	0.7076	&	4	&	0.0012	&	0.4888	&	0.3220	\\
      66	&	0.0013	&	0.0863	&	0.0875	&	0.7076	&	5	&	0.0007	&	0.3010	&	0.2450	\\
      66	&	0.0013	&	0.0863	&	0.0875	&	0.7076	&	6	&	0.0112	&	1.0000	&	0.9030	\\\hline

      68	&	0.0007	&	0.0455	&	0.0455	&	0.5499	&	1	&	0.0336	&	1.0000	&	1.0000	\\
      68	&	0.0007	&	0.0455	&	0.0455	&	0.5499	&	2	&	0.0387	&	1.0000	&	1.0000	\\
      68	&	0.0007	&	0.0455	&	0.0455	&	0.5499	&	3	&	0.0132	&	1.0000	&	1.0000	\\
      68	&	0.0007	&	0.0455	&	0.0455	&	0.5499	&	4	&	0.0119	&	1.0000	&	1.0000	\\
      68	&	0.0007	&	0.0455	&	0.0455	&	0.5499	&	5	&	0.0160	&	1.0000	&	1.0000	\\
      68	&	0.0007	&	0.0455	&	0.0455	&	0.5499	&	6	&	0.0052	&	1.0000	&	1.0000	\\
      68	&	0.0007	&	0.0455	&	0.0455	&	0.5499	&	7	&	0.0002	&	0.0654	&	0.0630	\\
      \hline
    \end{tabular}}

\end{table}

\section*{Discussion}
We introduced a conceptual idea for spatial high-dimensional data: dimension reduction in two steps, collapsing and clustering. These two steps are performed separately: fast collapsing on all array features and more rigorous, model-based clustering on the resulting regions, much fewer in number.

\subsection*{Conditional versus unconditional null-hypotheses}
The examples presented here require only one data set to be available for the purpose of both clustering and testing. This implies that the null-hypotheses are conditional on the clusters. However, the methodology developed here also applies to the unconditional setting, where the inference is performed on samples independent of those used for clustering. First, note that since the clustering algorithm always separates chromosomes, the conditional and unconditional nulls are likely to be very accordant on the chromosome level.

In the conditional setting, we emphasize that
the FWER on clusters estimated from the same data is well-defined, albeit from conditional, and hence random, null hypotheses (because the clusters are random). This point of view was investigated by \cite{Pacifico2004} in the context of a clustered version of FDR.
However, as opposed to  approach of \cite{Pacifico2004} using $p$-value-based clustering, we argue that the conditional approach introduced here clearly delineates the information in the data which is used to make clusters from the information used to perform testing on the clusters, because the clusters are permutation invariant, so that we obtain an interpretable procedure on the permutation probability space as well.

We acknowledge that the conditional setting is not fully in accordance with the traditional hypothesis testing setting, which is why one might prefer the unconditional setting in some cases. 
Reasons why we emphasize the conditional setting are the following. Tumor array CGH is still very expensive, and, more importantly, for many clinical studies it is not practically feasible to obtain good quality tumor material for a large amount of samples. Moreover, cancer is a very heterogeneous disease, and hence the power to detect differences between groups for particular genomic locations is likely to be small for small sample sizes. For example, some aberrations occur in only 5\% of the entire population, but it may still be relevant to detect such an aberration when (most of) it 5\% would belong to one group. Therefore, one often prefers to use as many samples as possible in the testing phase, and not `sacrifice' samples for the purpose of clustering. However, the unconditional setting may be particularly attractive when aCGH data ($X$) is externally available, but the response ($Y$) is not. In such a case, the external data can anyhow not be used in the testing phase. Our software easily applies to such a setting as well.

Finally,
we studied to what extent the results of the conditional and unconditional approaches differ for Data1. We repeatedly split this data set in two parts. The conditional approach uses the second part for both clustering and testing, whereas the unconditional approach uses the first part for clustering and the second for testing. We registered which regions are significant and which regions are a member of a significant cluster, at $\alpha=0.2$. Summarizing, on average 94.3\% (99.9\%) of the regions rejected (non-rejected) by the conditional approach were confirmed by the unconditional approach, whereas on average 93.2\% (99.8\%) of the regions that are member of a rejected (non-rejected) cluster as determined by the conditional approach were confirmed by the unconditional approach. Hence, it is comfortable to notice that the stability of the clustering implies a high agreement between the unconditional and conditional approach, whereas the latter uses only half of the data.

\subsection*{Other issues and conclusion}
A byproduct of our clustering method is the strength measure of association ($\gamma$) within a cluster. This helps in interpreting the results of the hierarchical multiple testing on clusters and regions: if $\gamma$ is relatively small for a significant cluster, one expects that the significance is driven by a few regions within the cluster, which should be reflected in the adjusted region-wise $p$-values. For large values of $\gamma$ one expects rather similar adjusted region-wise $p$-values.
Hence, to some extent $\gamma$ and the region-wise $p$-values may help in distinguishing genomic regions that are potentially causally related to the response and those that are just correlated with neighboring, more strongly associated regions. In any case, further biological validation, also at other molecular levels, such as mRNA or protein, is needed to decide which genomic DNA regions are really causally related to the response.

Some prefer the use of undiscretized rather than discretized array CGH data for testing association with clinical information. We believe our cluster-testing approach to be useful in this setting as well. The clustering would then group features that share the underlying discrete characteristics of the data, while the corresponding undiscretized data would be used to achieve (supposedly) more power in the permutation testing procedure.

For clusters that contain regions that are amplified (high copy number gain) rather than gained, one could apply our algorithm on a 4-digit code (-1,0,1,2) but this may be very time-consuming. When few losses (-1) are present in such a cluster (which is not unlikely given the presence of amplifications), one may locally re-define the 4-digit code to a 3-digit one where the losses and normals would be joined in one class. Another alternative solution is to keep the 4-digit code, but instead of using an extra parameter in model (\ref{eq:qem}) use the same parameter as for $x_i=1$ (gains), but now with $x_i=2$. This would give more weight to two consecutively amplified regions than for two consecutively gained regions. Double deletions may be dealt with in a similar way.

Our clustering algorithm suggests a development for DNA probe design in low dimensional platforms such as MLPA \citep{Schouten2002}. If strongly associated clusters could be verified in further studies, then one may consider to use only one MLPA probe per cluster. Such low dimensional platforms are usually less costly, more reproducible and more straightforward to analyze. Another application of our genomic clusters is the clustering of samples and prediction of clinical outcome. \cite{WieringenWecca} show that use of genomic regions rather than individual probes to cluster samples based on array CGH data enhances clustering performance. In prediction, the smaller number of clusters may result in simpler and more robust feature selection.

In summary, we developed a dedicated clustering algorithm which, in combination with permutation testing, allows a multiple resolution perspective on association of array CGH data with clinical information. With the introduction of extremely high-resolution data, f.e. obtained by massive parallel sequencing, the need for such methods will only increase.

 \section*{Acknowledgements} We thank Jelle Goeman and the Associate Editor for discussing the testing procedures with us and Wessel van Wieringen for providing code to create heatmaps.  The second author was partly supported by the French Agence Nationale de la Recherche (ANR) (references ANR-PARCIMONIE, ANR-09-JCJC-0027-01, ANR-09-JCJC-0101-01).

\section*{Appendix I: Software}
The methods discussed here, both for clustering and hierarchical testing, are implemented in the R-package dnaCplusT. The package contains example data sets, documentation on functions and parameters and the actual R-code.
For the implementation of the procedures, we used R-Bioconductor software packages {\tt RBGL, partitions} and {\tt multtest}.
The package is available from the last author's web site: http://www.few.vu.nl/$\sim$mavdwiel/dnaCplusT.html.

\section*{Appendix II: Validity of permutation test}
Here, we prove the validity of the permutation tests in our setting.
In this section, let us write $X_j=X_j^{\sbullet}$ for short.
Under the null
\begin{equation*}
  \mbox{$H_{0,j}$:\:\:\: $Y$ and $X_j$ are independent conditionally on $\widehat{\mathcal{A}}(X) = \mathcal{A}$},
\end{equation*}
the following proves that the distribution of $(Y,X_j)$ is equal to the distribution of $(Y^\sigma,X_j)$, conditionally on $\widehat{\mathcal{A}}(X)= \mathcal{A}$, for any deterministic permutation $\sigma$ of $\{1,...,n\}$
(where the superscript ``$\sigma$'' codes for a $\sigma$-column-wise permutation of the vector or of the matrix):
\begin{align*}
  \mathbf{P}( Y=y^\sigma, &X_j=x_j | \widehat{\mathcal{A}}(X)= \mathcal{A})\\ &=  \mathbf{P}( Y=y^\sigma | \widehat{\mathcal{A}}(X)= \mathcal{A}) \mathbf{P}(X_j=x_j | \widehat{\mathcal{A}}(X)= \mathcal{A})\\
  &=  \mathbf{P}( Y^\sigma=y^\sigma | \widehat{\mathcal{A}}(X^\sigma)= \mathcal{A}) \mathbf{P}(X_j=x_j | \widehat{\mathcal{A}}(X)= \mathcal{A})\\
  &=  \mathbf{P}( Y^\sigma=y^\sigma | \widehat{\mathcal{A}}(X)= \mathcal{A}) \mathbf{P}(X_j=x_j | \widehat{\mathcal{A}}(X)= \mathcal{A})\\
  &=  \mathbf{P}( Y=y | \widehat{\mathcal{A}}(X)= \mathcal{A}) \mathbf{P}(X_j=x_j | \widehat{\mathcal{A}}(X)= \mathcal{A})\\
  &=    \mathbf{P}( Y=y, X_j=x_j | \widehat{\mathcal{A}}(X)= \mathcal{A}).
\end{align*}

Similarly, we may prove that under the null
\begin{equation}\label{H0all}
  \mbox{$H_{0}^{\l}$:\:\:\: $Y$ and $(X_j)_{j\in A_\l}$ are independent conditionally on $\widehat{\mathcal{A}}(X) = \mathcal{A}$},
\end{equation}
the distribution of $(Y,(X_j)_{j\in A_\l})$ is equal to the distribution of $(Y^\sigma,(X_j)_{j\in A_\l})$, conditionally on $\widehat{\mathcal{A}}(X)= \mathcal{A}$, for any deterministic permutation $\sigma$ of $\{1,...,n\}$.\\

Null-hypothesis (\ref{H0all}) is sufficient for applying permutation and allows for FWER control using the inheritance principle. However, in order the sharpen FWER control using Shaffer's (\citeyear{Shaffer86}) improvement, we need to assume that (\ref{H0all}) is implied by
$$\bigcap_{j \in A_\l} H_{0,j},$$ because the intersection hypothesis contains the logical relationship between the region-wise null-hypotheses and the cluster hypothesis needed to apply this improvement. This assumption means that the dependency between $Y$ and the vector $(X_j)_{j\in A_\l}$ is fully described by dependencies between $Y$ and $X_j$'s, $j \in A_\l$. So for example $(Y \perp X_1) \land (Y \perp X_2) \Rightarrow Y \perp (X_1,X_2)$. Note the similarity between this assumption and one often made in step-wise regression (with random covariates), where an interaction term  $X_1X_2$ (which models dependency between $Y$ and $(X_1,X_2)$) is only considered once at least one of the main terms ($X_1$ or $X_2$) is present in the model.
% end change

\section*{Appendix III: Proof of FWER control}
Here, we provide a proof that the hierarchical multiple testing $\R$ controls the FWER both on the cluster and on the region levels. We consider the case where we include the Shaffer correction (the other case is similar). The proof was provided in \cite{Goeman2010}, but for a different type of hypothesis weighting. Our arguments are very similar.

The proof uses the so-called ``sequential rejection principle, presented by \cite{Goeman2009b} (see also \cite{Arlot2010}).
Our hierarchical rejection procedure $\mathcal{R}$, rejecting both clusters $A\in \mathcal{A}$ and regions $j\in A$, can be expressed as  the following sequential rejective procedure: $\R =\R_{k}$ where $\R_0=\emptyset$ and for all $i\geq 0$,
$$
\R_{i+1} = \R_i \cup \{A\in\A | p_A \leq \alpha_{\R_i}  \} \cup \bigcup_{A\in\A} \{j\in A | p_j\leq \alpha_{\mathcal{R}_i,A}\} ,
$$
and where $k$ is the first $i\geq 0$ for which $\R_{i+1}=\R_i$. In the above recursion relation, $p_A$ and $p_j$ are the $p$-values for cluster $A$ and region $j$, respectively, while the threshold on clusters is $\alpha_{\mathcal{R}}=\alpha/(K_{\mathcal{A}} - |E_{\mathcal{R}}|)$, with $E_{\mathcal{R}}=\{A\in \R\cap\A | A\subset \R\}$,  and  the threshold on regions is $\alpha_{\mathcal{R},A}=\alpha_{\mathcal{R}}\ind{A\in \A\cap \mathcal{R}}/(|A|-\max(1,|A\cap \R|))$. In the latter thresholds, we use the convention $1/0=+\infty$ and $0/0=0$.

We aim to establish that our procedure controls the hierarchical FWER at a pre-specified level $\alpha$. From the sequential rejection principle, the latter is true, as soon as both a ``monotonicity condition''  and a ``single step condition'' hold \citep{Goeman2009b}. The first condition is satisfied because $\alpha_{\mathcal{R}}$ and $\alpha_{\mathcal{R},A}$ are nondecreasing in $\R$. The second condition is satisfied if we have
\begin{equation}\sum_{A\in \A\cap \H_0}  \alpha_{ \H_0^c} + \sum_{A\in \A} \sum_{j\in A\cap \H_0}  \alpha_{\H_0^c,A}\leq \alpha ,\label{goal}\end{equation}
where we denoted by $\mathcal{H}_0$ the set of the true clusters and regions and by $\mathcal{H}_0^c$ its complementary. We now prove \eqref{goal}: %the left hand side of \eqref{goal} equals
we have
\begin{align*}
  \sum_{A\in \A} \sum_{j\in A\cap \H_0}  \alpha_{\H_0^c,A} &=   \alpha_{ \H_0^c} \sum_{A\in \A\cap \H_0^c} \frac{|A\cap \H_0|}{|A|-\max(1,|A\cap \H_0^c|)}.
\end{align*}
Now, the sum appearing in the right hand side of the above relation can be taken only over $A\in \A\cap \H_0^c$ such that $|A\cap \H_0|\neq 0$, that is over $A\in   \A\cap \H_0^c \backslash E_{\H_0^c}$.
Moreover, we may use the logical relation  between the hierarchical hypotheses saying that if a cluster is false then at least one of its regions is  false:  $A\in \A\cap \H_0^c$ implies $|A\cap \H_0|\leq |A|-1$. Combining these two facts, we obtain
\begin{align*}
  \sum_{A\in \A} \sum_{j\in A\cap \H_0}  \alpha_{\H_0^c,A} &=   \alpha_{ \H_0^c} \sum_{\A\cap \H_0^c \backslash E_{\H_0^c}} \frac{|A\cap \H_0|}{|A|-\max(1,|A\cap \H_0^c|)} \\
  &\leq \alpha_{ \H_0^c} \sum_{\A\cap \H_0^c \backslash E_{\H_0^c}} \frac{\min(|A\cap \H_0|,|A|-1)}{|A|-\max(1,|A\cap \H_0^c|)}\\
  & =  \alpha_{ \H_0^c} (|\A\cap \H_0^c| - |E_{\H_0^c}|)
\end{align*}
Thus, we have
\begin{align*}
  \sum_{A\in \A\cap \H_0}  \alpha_{ \H_0^c} + \sum_{A\in \A} \sum_{j\in A\cap \H_0}  \alpha_{\H_0^c,A}  &\leq \alpha_{ \H_0^c} (|\A\cap \H_0| + |\A\cap \H_0^c| - |E_{\H_0^c}|)\\
  &= \alpha_{ \H_0^c} (K_{\A}- |E_{\H_0^c}|) = \alpha,
\end{align*}
which proves \eqref{goal} and the required FWER control.
\bibliographystyle{natbib}
% \bibliography{../Reference/kikim}
\bibliography{KRW2010}

\end{document}